# Time to adjust: Improving replicability in experimental psychology by adjustment for evident selective inference

Yoav Zeevi[1,2], Sofi Astashenko[2], Liad Mudrik[1,3,4], Yoav Benjamini[1,2]

[1] The Sagol School of Neuroscience, Tel Aviv University, Tel Aviv, 39040, Israel

[2] Department of Statistics and Operations Research, Tel Aviv University, Tel Aviv, 39040, Israel

[3] School of Psychological Sciences, Tel Aviv University, Tel Aviv, 39040, Israel

[4] Canadian Institute for Advanced Research (CIFAR), Brain, Mind, and Consciousness Program, Toronto, ON, Canada

**Abstract**

The field of psychological sciences has been grappling with the replicability crisis. Various issues have been identified as potential sources of this problem. We bring to light a potential source that has largely been overlooked and demonstrate its significant contribution to the problem: the practice of multiple comparisons. We analyzed 88 papers from the Reproducibility Project in Psychology and found that multiple results are commonly reported in a single paper, ranging from 4 to 730 (M=77.7), without multiple comparison adjustments. We retroactively applied such an adjustment using a hierarchical FDR controlling procedure (TreeBH; Bogomolov et al., 2021). 21 of 88 results were deemed insignificant after adjustment. Twenty of these 21 results indeed failed to replicate, constituting over a third of the non-replicable findings, while maintaining 97% power. We propose that this should become a common practice as an essential means to increase replicability in experimental psychology.

**Research Transparency Statement**

All materials, dataset, and codes utilized and analysed during this study is available in the supplementary materials accompanying this manuscript. This study was not preregistered.

## Introduction

Recent years have witnessed a dramatic increase in the number of papers addressing the problem of replicability in many different scientific fields (Fanelli, 2009; Peng, 2011; Prinz et al., 2011; Gelman & Loken, 2013; Achenbach, 2015; Baker, 2016; Lithgow et al., 2017; Frank et al., 2017; Nosek & Errington, 2017). Psychology is one of the first research fields where the replicability problem has received significant attention (Smith, 1970). Later on, a series of high-profile papers demonstrated the low replicability rates in the field (Klein et al., 2014, 2018; Open Science Collaboration, 2015; Ebersole et al., 2016; Camerer et al., 2016, 2018); collapsed across all these papers, only 90 of 190 replication attempts were indeed replicated. Thus, it is no wonder that the issue is often called the replicability crisis (Methner et al., 2023).

This proclaimed threat to the credibility and reliability of the field has led to an essential movement of self-correction, whereby the potential sources of this problem were discussed. One such critical source, which has been examined both outside (Xie et al., 2021) and within the field of psychology (Gelman, 2013), is *selective inference*: focusing statistical inference on some findings that turned out to be interesting after viewing the data (Benjamini, 2020). Such selective inference characterizes both the issue of publication bias (also termed the file-drawer problem, Rosenthal, 1979) and that of General Questionable Research practices (QRPs; Ravn & Sørensen, 2021). QRPs relate to the unreported flexibilities researchers take with data analysis before casting their results for publication. It is a major playground for selective inference, including p-hacking (Simmons et al., 2011), hypothesizing after the results are known (HARKing, Kerr 1998), trying multiple statistical tests and reporting only the best ones, optional stopping of data collection, manipulation of outliers, malicious exclusion of participants, etc., - all in the effort to obtain significant results (John et al., 2012; Button et al., 2013).

Importantly, these are all instances of selective inference that is *not evident in the paper*; since the authors do not report the different analyses that were conducted, but only the successful ones, it is hard to estimate the actual reliability of the reported results. However, we claim that the same problem also applies to selective inference that *is evident in the paper*, demonstrated below to have a substantial detrimental effect on the reproducibility of a given result, which has been mostly overlooked so far. This significant source for the replicability crisis can be easily addressed, leading to a more robust and reliable corpus of knowledge in experimental psychology.

*Evident selective inference* pertains to a situation whereby some results out of the many reported in the paper are highlighted as important, without adjusting their significance for the multiple results (i.e., comparisons) from which they were selected to be highlighted. This is very prevalent in psychology and cognitive science (and other fields; see Benjamini 2020). The results that are not highlighted are either reported elsewhere (i.e., in the supplementary) or are not reported at all but can be inferred (for example, based on the experimental design). Unsurprisingly, the highlighted results are commonly the statistically significant ones. Such evident selective inference is unavoidable in the current 'industrialized' way science is conducted, where new technologies for data generation (computerized tracking system, fMRI, genomics, etc.), data storage, and computing software increase the number of statistical tests and inferences conducted in a single study. Moreover, in many ways, the evident selective inference is an inherent aspect of writing manuscripts: one must highlight the important findings, summarize them for the reader, and discuss their importance. We support this view but argue that doing that without adjustment for the selection increases the risk of making false-positive discoveries (Simmons et al., 2011). Unless adjusted for selection, the selected p-values and confidence intervals lose their probabilistic meaning (Mayo, 2018).

To explain, let us focus on one example: a paper by Schnall and colleagues (2008). The study had 140 evident comparisons (two priming methods, ten emotions with seven average scores for each emotion; full description can be found in the supplementary materials). The authors reported only six p-values, all statistically significant without multiple comparisons adjustment. Given the 140 evident comparisons, the expected number of statistically significant p-values at 5% (even if there is no effect at all) was 7, ($= 140 \cdot 0.05$), so doubt should be cast on the statistical significance of these six results. Moreover, even if we assume that there were a few real effects, given the expected seven false effects, it would still be reasonable to assume that the rate of false claims among the claims is higher than desired. Indeed, the study's findings failed to be replicated twice, once with the same sample size (Open Science Collaboration, 2015) and the other with almost three times as many participants (N=126; Johnson et al., 2014). Let us emphasize: to the best of our judgment, this is not a case of non-evident selective inference. All the information about the different experimental conditions appears in the paper, making it possible to infer the number of comparisons that were likely performed.

Accordingly, the goals of the current paper are fourfold. First, to examine how frequent evident selective inference is in the field (i.e., assess the level of results-multiplicity in psychological papers). Second, to check how often researchers adjust for such evident selective inference. Third, to provide a method that allows appropriate adjustment for selective inference (TreeBH; Bogomolov et al., 2021; see *Adjusting for evident selective inference in the original publications using the TreeBH method* below), while importantly preserving power (i.e., the ability to detect a true effect). Finally, to demonstrate the extent to which applying the suggested method can improve the replicability of the original study's results. Thus, this work not only showcases a prominent problem in the field, but also suggests a simple solution that could mitigate its effect and increase replicability.

## Methods

### The database: The Reproducibility Project in Psychology (RPP)

We evaluated all studies (N=100) included in the Reproducibility Project in Psychology (RPP, Open Science Collaboration, 2015). In this study, multiple research groups attempted to replicate one meaningful and statistically significant result from each of 100 papers published in 2008 in Psychological Science (PSCI), Journal of Personality and Social Psychology (JPSP), and Journal of Experimental Psychology: Learning, Memory, and Cognition (JEP: LMC). Notably, some criticisms have been leveled at the RPP regarding the extent to which the original experiments' methods were followed (Stroebe, 2016; Gilbert et al., 2016). Still, our rationale here is that this database provides, to the very least, a rough estimation of the replicability and robustness of these 100 original papers.

This database uniquely suits our point, as it includes both the original works and their later replicability results. Specifically, we first examined the prevalence of evident selective inference in the original RPP papers, and then adjusted for selection where needed. We utilized the False Discovery Rate (FDR) approach (Benjamini & Hochberg, 1995), applying a relatively new hierarchical FDR methodology (TreeBH; see Bogomolov et al. 2021) that exploits the hierarchical structures common in experimental psychological studies to gain power and relevance. Then, we reexamined the replicability results in light of the adjustment findings and asked how many replicated/non-replicated findings were deemed significant/insignificant following this adjustment. We hypothesized that many non-replication results would be insignificant following the adjustment. This would suggest that adjusting for evident selective inference is a potent method for reducing the proportion of false discoveries, thereby decreasing the probability of non-replicable findings being published to begin with.

Out of the 100 replication attempts, we excluded twelve studies for the following reasons: Six studies had an original result of $p > 0.05$ upon the replicators' recalculations (although initially reported as statistically significant results). Three studies originally reported non-significant results (and the replications accordingly attempted to establish non-significance, which is less relevant to our purposes). One study reported a significant result, but there was not enough information in the original paper to recalculate the p-value. One more replication attempt did not focus on obtaining a significant result and instead aimed at replicating the value of Pearson's r. Finally, one original result was tested in two replication attempts. The 88 remaining studies were contrasted with their replication attempts.

## Analysis

All analyses were performed using R Core Team (2021). The codes and data files are openly shared and can be found in the supplementary materials.

**Assessing the number of evident selective inference (i.e., level of multiplicity)**

The number of comparisons in each paper was assessed by two authors (YZ and SE), or three (YZ, SE and YB) when in doubt. We estimated the number of evident comparisons based on the reported results of each original publication.

To illustrate the estimation process, we will use an example from the database: a paper by Goschke & Dreisbach (2008), where participants had to notice rare prospective memory cues (Figure 1). Five 3-way ANOVAs were designed to support similar hypotheses with different dependent variables. For each of the 3-way ANOVAs, seven p-values were calculated: three main effects, three 2-way interactions, and one 3-way interaction. Taken together, the study included 35 comparisons. The conclusion was that rare prospective memory cues were overlooked more often on non-compatible trials than compatible trials and on task relevance rather than irrelevance. This was based on a 2-way interaction between the two variables (compatibility and task relevance), $F(1, 38) = 6.21$, $p=0.017$. This result was the target for replication in the RPP.

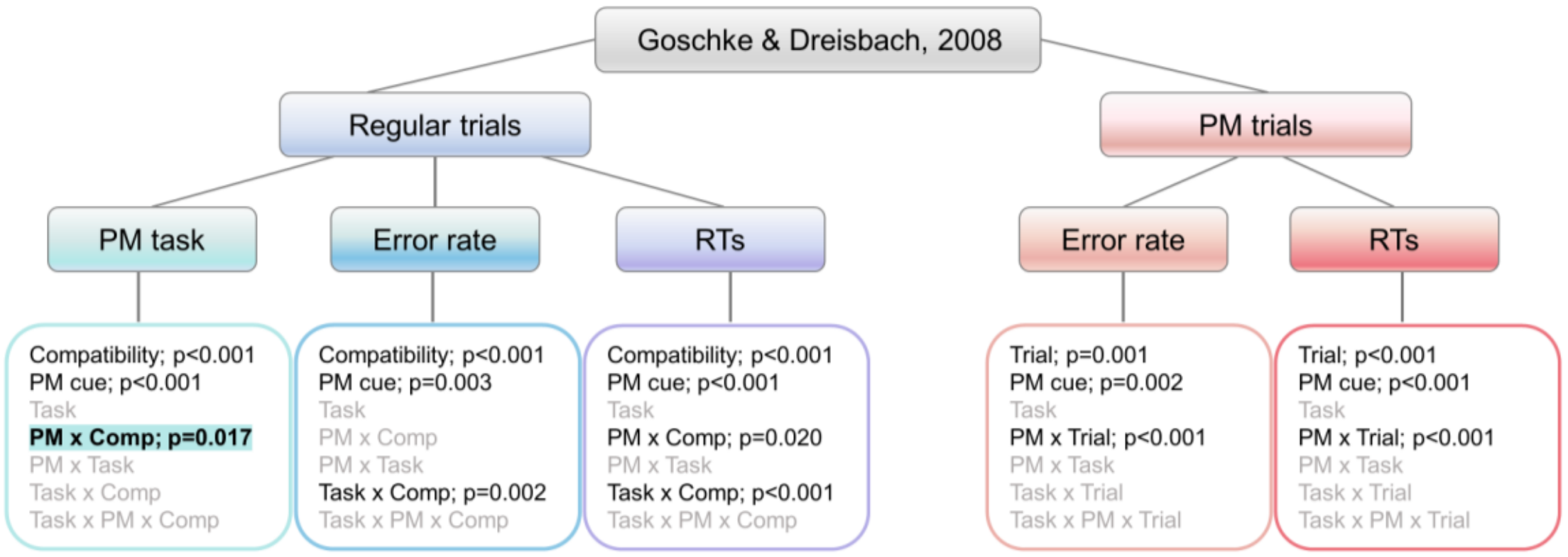

*Figure 1: Detailed analysis of five 3-way ANOVAs from Goschke & Dreisbach (2008) comparing error rates and response times (RTs) in regular and prospective memory (PM) trials. The p-values are indicated for statistically significant results (at 0.05), with the value highlighted in blue targeted for replication. Greyed-out represents nonsignificant results.*

## Adjusting for evident selective inference in the original publications using the TreeBH method

The TreeBH method is based on the widely used FDR criterion (Benjamini & Hochberg, 1995) for multiple inference problems. The false discovery proportion is the proportion of type-I errors among the discoveries made (0 if none is made), and the FDR is its expected value. In the context of testing, controlling the FDR at the 0.05 level maintains that, at most, 5% of results deemed "statistically significant" will be false positives on average. Regular unadjusted (i.e., uncorrected) testing fails to guarantee it, so Benjamini & Hochberg have introduced the BH procedure for controlling the FDR at a desired level q (also known as α, usually 0.05. For full description, see supplementary materials). If many true results are evident (i.e., many p-values are smaller than q), the BH method will require a minimal adjustment, if at all, on the p-values; if merely one exists, the BH method will adjust it as the traditional

Bonferroni does. The rationale here is that while having two false discoveries out of 50 is bearable, it is unbearable to have two false ones out of four. With the many variations introduced later, the FDR approach and the BH procedure have gained extensive usage in many scientific fields (Zucker et al., 2019; Korisky et al., 2021; Pudjihartono et al., 2022; Zhang et al., 2022; Amir et al., 2023; Heumos et al., 2023).

However, the BH approach considers all hypotheses similarly. In common practice, hypotheses are nested hierarchically (see Figure 1). That is, in a paper reporting several experiments, each experiment has its own hypotheses. Moreover, within each experiment, some hypotheses are conditioned upon finding a significant result at a higher level (e.g., one should only examine simple effects if the interaction is statistically significant).

The TreeBH method was proposed to exploit this hierarchical structure to increase the power of the adjustment method while still properly controlling FDR (Bogomolov et al., 2021). The method controls FDR at each level separately while making the necessary adjustments. Thus, it uses the FDR in a way that follows the structure of the inference. Here, we provide the rationale, illustrating how this method should be applied in future studies. To better explain the process, we use Goschke and Dreisbach's (2008, figure 1) example.

The TreeBH method is composed of three stages: building the analysis tree, calculating and adjusting the p-values, and then adjusting the significance threshold (i.e., alpha) to determine which comparisons survive the adjustement.

**Stage 1: Building the analysis tree**

First, researchers should *identify the hierarchical structure* of the hypotheses and construct the study tree. This stage should be conducted at the design stage, where the research hypotheses are spelled out (see again Figure 1, where the nested structure of the comparisons is described for the Goschke and Dreisbach, 2008 study). The guiding principle should be to identify families of hypotheses related in meaning and their inter-dependencies (i.e., which hypotheses are driven by, or depend on, others). Notably, the tree should only include the comparisons of interest (i.e., all the comparisons that would make the paper publishable if found significant). For example, in a 3-way ANOVA, if the researchers know at the design stage that they are not interested in one or more of the main effects or one or more of the 2-way interactions, they can choose not to include them in the tree, thereby forgoing the need to adjust for these comparisons. However, if the tree is only constructed at the analysis stage, all should be considered jointly.

**Stage 2: Calculating and adjusting the p-values**

After collecting the data, one should run the analyses, *calculate the p-values* for every hypothesis, and place them on the tree. The overall goal is to have a p-value for each node in the tree. Notably however, the highest levels of the tree might be ANOVAs, like in Goschke & Dreisbach (2008, Figure 1), or even an entire experiment (e.g., when the manuscript includes several experiments, each experiment might be a node, out of which several analyses nodes emerge). How can one determine the p-value of the ANOVA node or of an entire experiment? Different approaches can be taken, but here, we propose to use the Simes test (Simes, 1986; See Figure 2), which increases the power of the TreeBH procedure and is proven to offer FDR control.

The TreeBH procedure is based on the assumption that the lower-level 'children' effects represent in a more refined way the same phenomenon as the higher-level 'parent' one (i.e., if a significant 'parent' effect is found – like a main effect, it should imply that at least one 'child' being significant at the same $\alpha$-level– for example, a simple effect). Accordingly, the Simes test asks if there is at least one significant result within the family, which means that using the Simes test ensures the TreeBH assumption is met.

Thus, a process of assigning p-values ensues, starting from the lowest levels of the tree and moving upwards. At each step in this stage, the classical *BH adjustment is applied within* the family of comparisons, sharing the same parent at one level above, *producing BH-adjusted p-values.* Then, the smallest BH-adjusted p-value within that family is assigned to its parent.[1]

---

[1] Note that while it is possible to replace the p-value of a main effect with the Simes test of the post-hoc comparisons, this cannot be done for an interaction effect, as the test's assumption is not met; it is not sufficient for one simple effect to be significant for an interaction to be found; it has to be significantly different from the other simple effect (i.e., the difference between differences). Thus, one cannot replace the p-value of the interaction effect with the Simes test of the simple effects. In general, Simes should not be used where the children's hypotheses do not represent the refined parent's hypothesis.

Returning to the example study by Goschke and Dreisbach (Figure 02), in step 1, we use the BH method to adjust the seven p-values within each of the 3-way ANOVAs. Then, the Simes test is applied, such that the smallest adjusted p-value within each 3-way ANOVA cluster is propagated up as the representative value for that ANOVA. As a result, each ANOVA has its own p-value, equivalent to the smallest p-value of the main effects within each of these ANOVAs.

In step 2, the three ANOVAs p-values are adjusted using the BH method, akin to the adjustment done in Step 1 within each ANOVA. Again, the smallest of these adjusted values is moved up the hierarchy. This p-value represents the entire left branch (regular trails), and the same procedure is applied to the right branch (PM trials). Thus, each of the two analysis branches – one for all analyses applied on regular trials and the other for all analyses applied on PM trials – now have their own p-values.

In step 3, the same procedure is applied, yet now, on Level 2, the two p-values obtained in step 2 are adjusted, and the lowest of them is selected to move up in the hierarchy. This determines a single p-value that represents the study's overall findings.

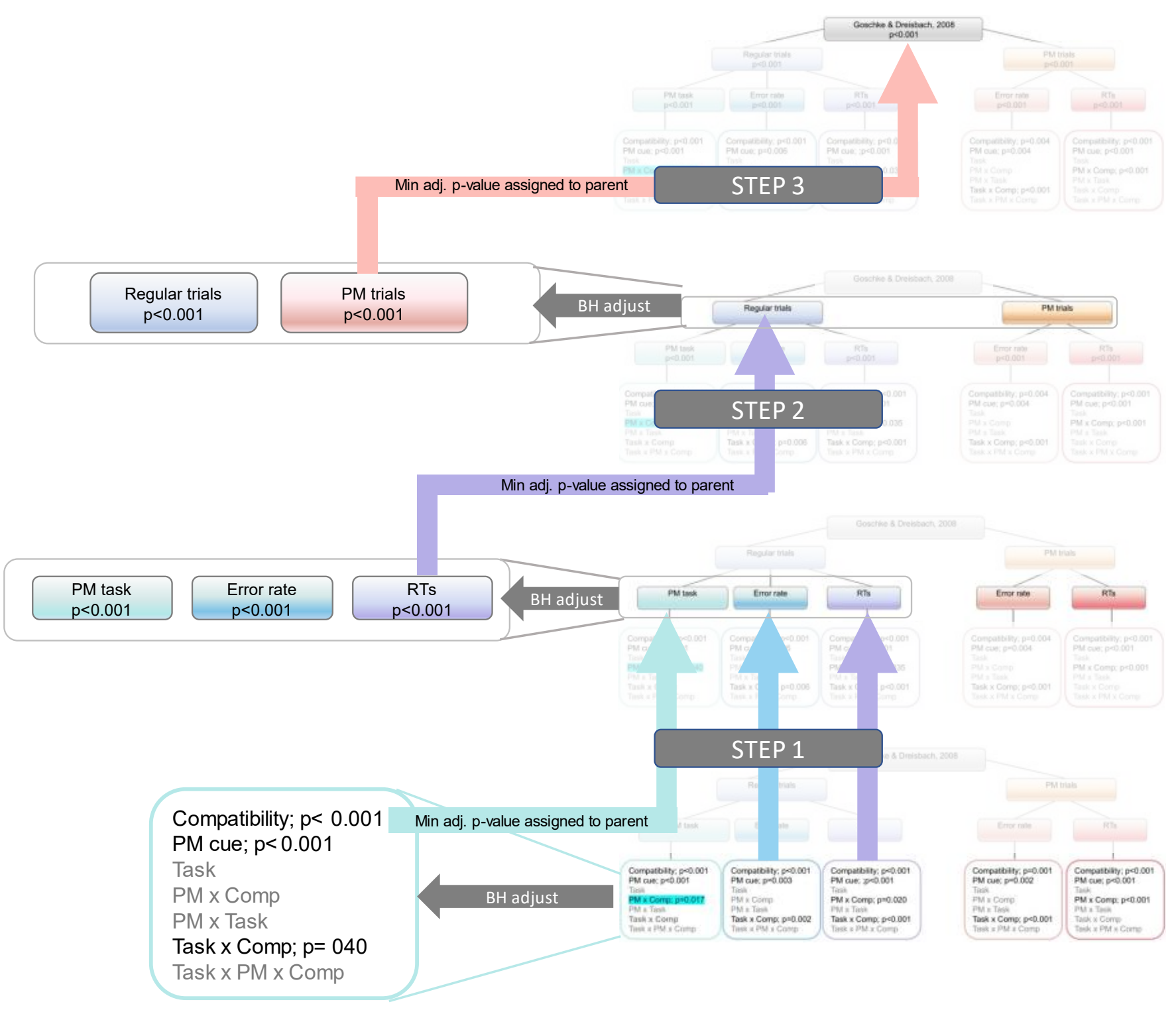

Figure 2 : Flowchart illustration of the Simes test's hierarchical application in the example study by Goschke and Dreisbach. The Full analysis can be found in the supplementary material. In Step 1, the BH method is used to adjust the seven p-values within each of the 3-way ANOVAs. The smallest adjusted p-value within each 3-way ANOVA cluster is propagated up as the representative value for that ANOVA. In Step 2, the three ANOVA Simes p-values are adjusted using the BH method. The smallest adjusted p-value is selected and moved up the hierarchy, representing the entire left or right branch of the analysis. Finally, in Step 3, the two p-values from step 2, representing regular and PM trials, are adjusted again. The lowest adjusted p-value is chosen to represent the study's overall findings.

### Stage 3: hierarchical testing with alpha adjustment

After all p-values are calculated and adjusted, we begin the 3rd stage: adjusting the alpha. The adjustment takes into account the number of significant comparisons at the upper level. Note that this process moves downwards, from top to bottom. Thus, specifically for Level 2, the adjusted alpha would be calculated in the following manner:

$$\alpha_{adjusted} = \alpha \cdot \frac{\text{\#significant p values in Level 1}}{\text{\#p values in Level 1}}$$

This procedure is then repeated for the ensuing levels. For Level L+1, the adjusted alpha from Level L is multiplied by the ratio of significant results in Level L to the number of hypotheses at that level. That way, each level in the hierarchy affects its subsequent dependent comparisons:

$$\alpha_{adjusted,L+1\ level} = \alpha_{adjusted,L\ level} \cdot \frac{\text{\#significant p values in Level L}}{\text{\#p values in Level L}}$$

Figure 3 provides a hypothetical example of the alpha adjustment in the Tree BH method. The top node represents the first hypothesis ($H_{11}$) with an alpha level ($\alpha_1$) set as α. The full circle at $H_{11}$ signifies a significant result, prompting further testing of three subsidiary “blue family” hypotheses ($H_{21}$, $H_{22}$, $H_{23}$), with an adjusted alpha level ($\alpha_2$) still the same as $\alpha_1$.

Two of the three hypotheses are statistically significant, which leads to further testing within the dark green ($H_{3L1}$, $H_{3L2}$) and purple ($H_{3R1}$, $H_{3R2}$, $H_{3R3}$) families, with $\alpha_3$ adjusted to be two-thirds of $\alpha_2$. The fourth level presents the light green family ($H_{4L1}$, $H_{4L2}$, $H_{4L3}$), with alpha level $\alpha_{4L}$ adjusted to be half of $\alpha_3$ as only half of the dark green hypotheses were statistically significant. On the right side of the fourth level, the pink family is tested against $\alpha_{4R}$, which is one-third of $\alpha_3$, as there is only one statistically significant result in the purple family. Accordingly, in this process, the critical significant threshold alpha becomes smaller (i.e., harder to pass), and the more insignificant comparisons are present in the previous levels leading to it. In contrast, if all comparisons are significant, there is no cost to having multiple ones. Accordingly, researchers do not "pay a price" for having multiple comparisons as long as they yield significant results.

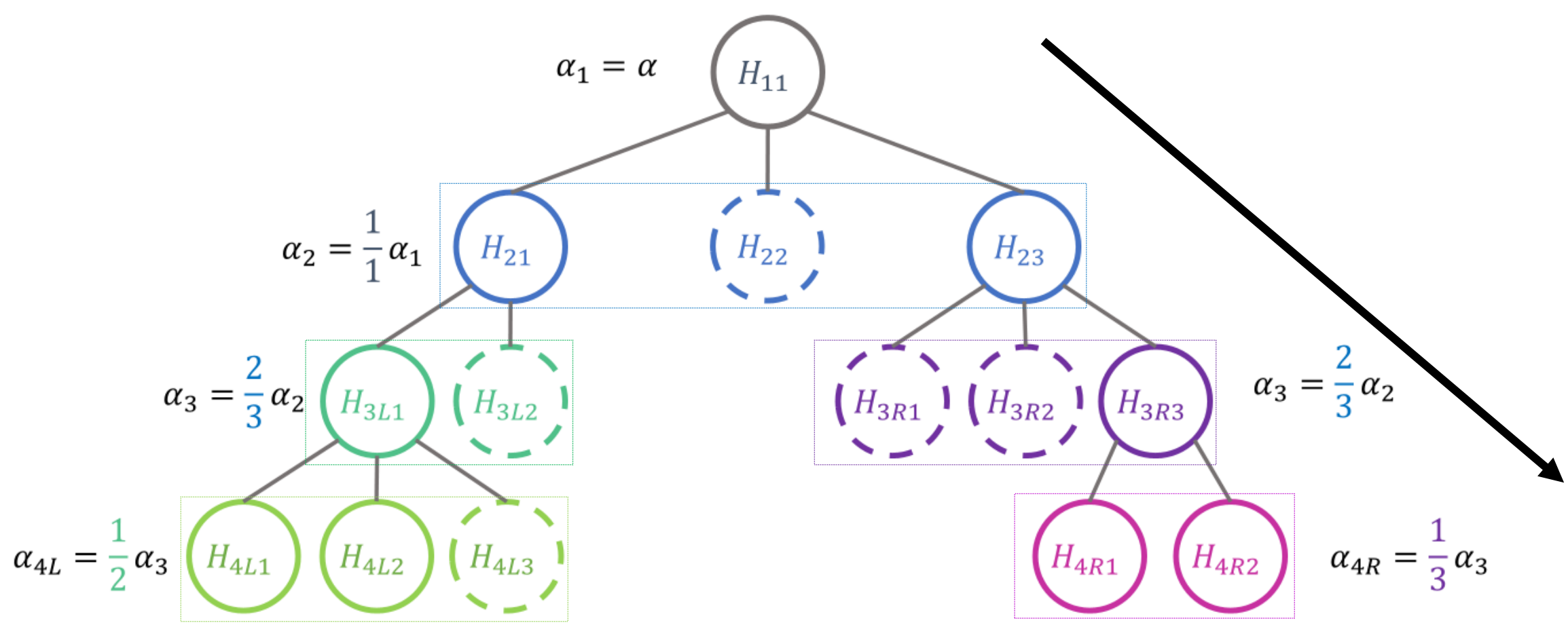

Figure 3: Hierarchical representation of hypothesis testing using the TreeBH method, showcasing the stepwise adjustment of α-levels across different hypothesis families. Solid circles indicate statistically significant hypotheses at the adjusted α-level, while dashed circles denote non-significant results. Each squared color grouping represents a distinct family, with the α-level adjustments displayed for successive testing stages. Alpha is multiplied at each stage based on the ratio of significant to all comparisons in the previous level, illustrated here by coloring the ratio in the color of the relevant level.

Applied to the Goschke and Dreisbach data, the p-value of the entire paper is statistically significant at 0.05 (1/1), so no alpha adjustment is needed for the second level. Both branches at Level 2 are statistically significant at 0.05, so two significant out of two implies using 0.05(2/2), so no alpha adjustment is needed for the third level either. The p-values from the three 3-way ANOVAs on the left are all statistically significant (3/3), so again, the alpha remains unaltered at Level 4. The final step involves comparing the adjusted p-value of the replication target (0.04) to the adjusted alpha at Level 4 (0.05). The result is deemed statistically significant at the 5% level. Note that this would not have been the case using other post-hoc correction methods (e.g., Bonferroni; Rice, 1989): such methods would simply divide the α=0.05 by 35, making the key finding insignificant. This shows that the TreeBH method takes a more lenient approach, which maintains power on the one hand yet minimizes false discoveries on the other hand.

**Assessing replicability with respect to adjusting for evident selective inference**

We applied the TreeBH method described above to the 88 included studies wherever possible. In studies without hierarchical structure, we used the regular BH adjustment method. To reduce redundancies in the text, we henceforth call the adjustments TreeBH even if it was only the regular BH. Our goal was to assess the replicability of the results as they were initially reported and as they were found following multiple comparison adjustments. Replicability was assessed by testing if the results were also found to be statistically significant in the replication study, using the same prescribed level while being consistent in direction. We note here that other criteria for replicability have been offered (Valentine et al., 2011; van Aert & van Assen, 2017, 2018; Hung & Fithian, 2019). However, we have used the 0.05 criterion only to assess the impact of selective inference on the replicability of the original study results. We were not interested in determining which results were indeed well established. Hence, in this work, we avoid determining which criterion for replicability is best and adopt the criterion used in this replication project (RPP, Open Science Collaboration, 2015).

## Results

### Number of evident selective inference (i.e., level of multiplicity) in the RPP papers

For each paper in the RPP, we estimated the number of evident tests in the paper (see Methods above). This number ranged from a minimum of 5 to a maximum of 730 (M= 77.7, SD=95.7). The number of evident tests in papers from Cognitive Psychology (M=61, Md=26.5, SD=72, N=26) was numerically lower than in papers from Social Psychology (M=100, Md=69, SD=123, N=44), yet this difference was insignificant following the TreeBH adjustment ($t_{35} = 1.95,\ p = 0.057,\ p_{adj} = 0.114$; Figure 4).

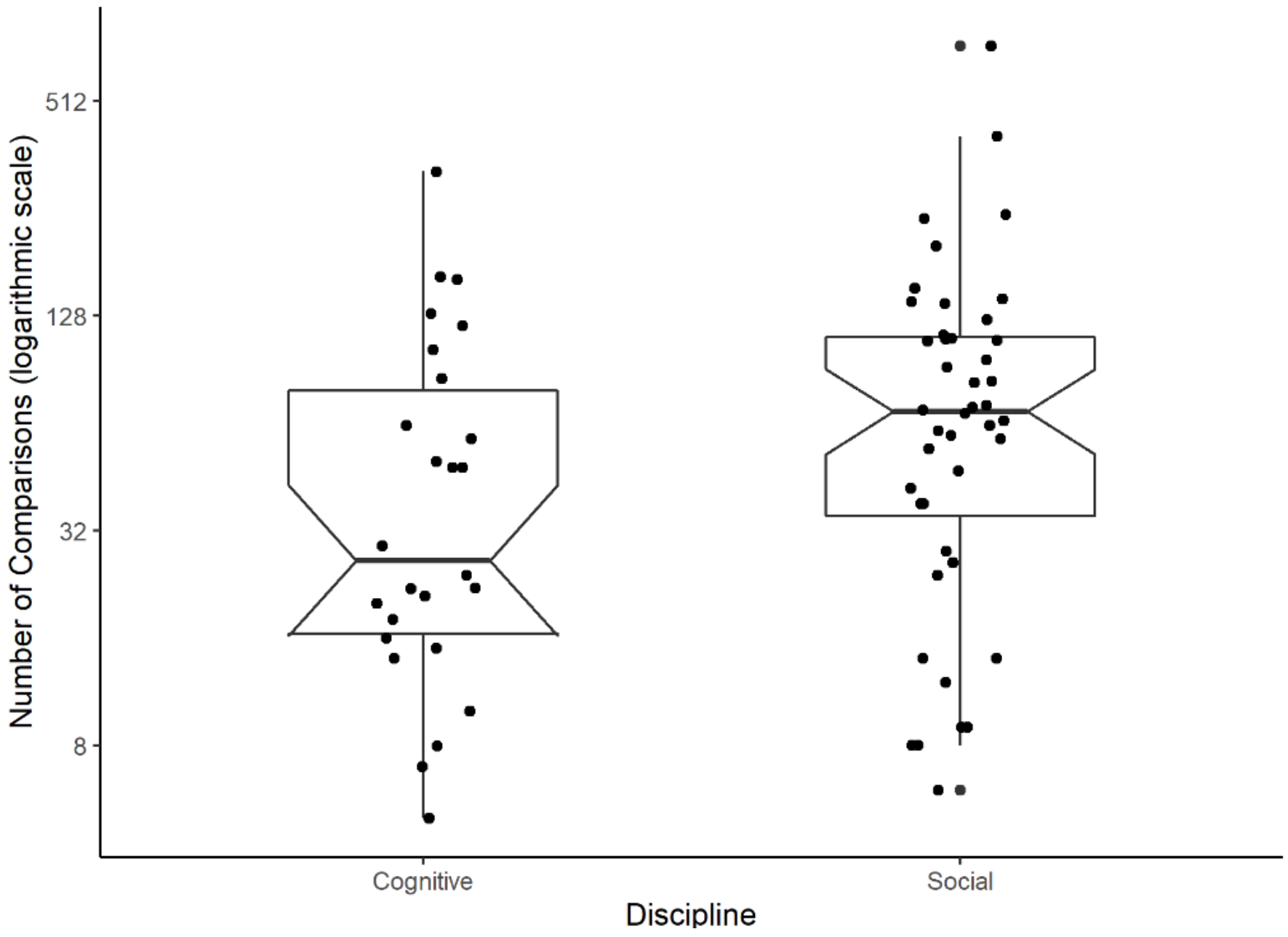

Figure 4: Notched boxplot of the number of evident comparisons per paper (log scale) in Cognitive and Social Psychology. The fact that the notches do not overlap indicates a statistically significant difference among the medians displayed by the central bar. Each point represents a single study.

### Adjusting for evident selective inference using the TreeBH method

In the original papers included in the RPP database, selective inference received very little attention. In most papers, there were few to no multiple comparisons adjustments or any text dedicated to discussing this problem. Only eight papers reported any selection adjustments. Not a single paper was adjusted for selection appropriately. That is, no paper included all its statistical tests in a comprehensive, selective inference adjustment framework. The somewhat surprising result here is that the papers that did make some, albeit partial, adjustment for selective inference had fewer comparisons (M=40, Md=24, SD=34.5, N=8) than those who did not adjust at all (M=83, Md=58, SD=102, N=85), but again, this difference was not statistically significant ($t_{10} = 1.55$, $p = 0.15$, $p_{adj} = 0.15$).

We applied the TreeBH approaches to all 88 papers included. The results per paper are depicted in Figure 5 (see details in Supplementary Material), and their summary is provided in Table 1. Our key question was how many non-replicated results could be detected as non-significant using the TreeBH and how many replicable ones could be detected as significant. We found that 20 out of the 56 results that failed to be replicated were deemed insignificant by the TreeBH method. This narrows down the error by 35%. That is, had the original studies applied the TreeBH method to their data, 35% of the results reported as significant, which ended up being non-replicable, would have been considered non-significant to begin with. Although still not guaranteeing replicability, the TreeBH method thus substantially minimizes non-replicability. One might be concerned that this benefit comes at a cost of reduced power; the results clearly suggest otherwise. Strikingly, 97% of the replicable results (31 out of 32 results) were indeed found to be significant using our method. Assuming that replicability implies a genuine effect, the vast majority of genuine effects were indeed detected by this method. Under the same rationale, the method has a low rate of false negative results; only 1 out of 21 results (5%) was replicated, although found insignificant using the TreeBH method (Table 1). Using the method also increased replicability, although to a limited extent, from 36% (32 out of 88, according to the original RPP) to 46% (31 out of 67). Thus, although a positive result in the TreeBH method does not guarantee replicability, using the method increases the chances that a statistically significant result will indeed be replicated without losing replicable results.

| **Selective inference adjustment** | | **Replication's p-value** | | **Total** |
|---|---|---|---|---|
| | | **Successfully replicated ($p \leq .05$)** | **Failed to replicate ($p > .05$)** | |
| **Original result after TreeBH adjustment** | $p_{adj} \leq .05$ | 31 | 36 | 67 |
| | $p_{adj} > .05$ | 1 | 20 | 21 |
| **Total** | | 32 | 56 | 88 |

Table 1: Replication before and after FDR adjustment for selective inference.
$\chi^2(df = 1, N = 88) = 11.9, p < 0.0001, p_{adj} < 0.0001.$

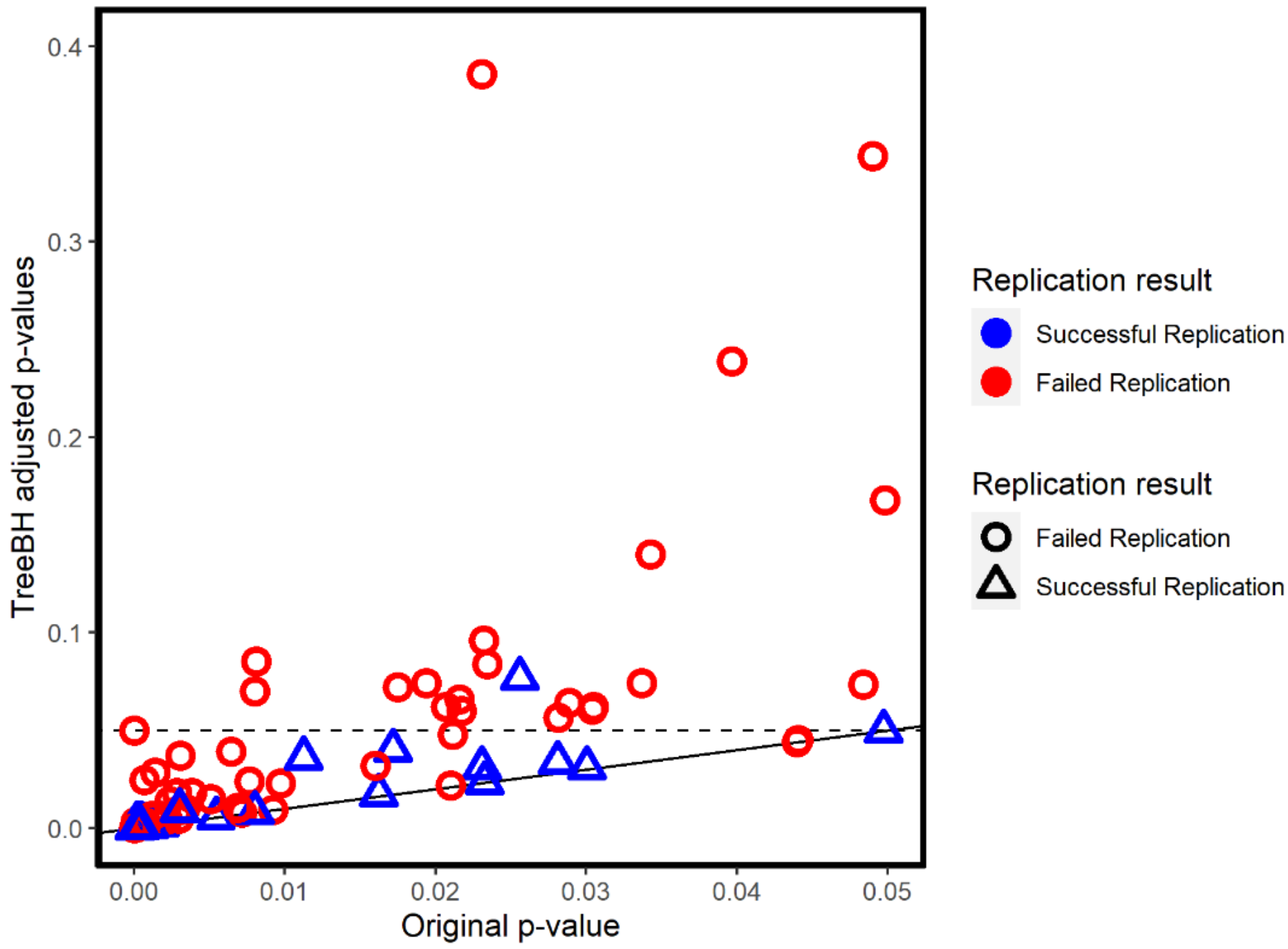

Figure 5: Scatter plots of the TreeBH adjusted versus the original p-values. Red circles depict failed replications, while blue triangles represent successful ones. The dashed line depicts the significance threshold for the TreeBH procedure. Since all included results were originally significant, no significance threshold line for the original p-values is depicted (and they range between 0 and 0.05). The solid line represents no change in p-value due to multiple comparisons. The further away a dot is from the solid line, the greater the estimated effect of multiple comparisons on the obtained p-value. As the figure shows, many failed replications received a much higher p-value (i.e., deemed less significant) following the Tree-BH method, while replicated ones were mostly unaffected. This suggests that papers that failed to replicate indeed suffered from a more severe problem of multiple comparisons.

In summary, we showed that "statistically significant" results that were deemed not significant after appropriate correction for multiple comparisons rarely got replicated in the tested database (1 in 21, which, incidentally, is the type I error rate). Moreover, by adjusting for FDR, replicability can be enhanced (from 36% to 46%) without giving up much power (one out of the 32 replicated results were not significant after the adjustment, Table 1). Notably, two results found to be statistically significant after TreeBH adjustment and did not replicate in the RPP were later criticized and replicated under revised protocols (Ebersole et al., 2020), implying that the conclusions from Table 1 are even more robust, with a 49% replication rate.

One could claim, however, that our adjustment's success in increasing replicability stems from its tendency to exclude weak results: adjusting for multiple comparisons effectively filters out weaker results, and these results are less likely to replicate. Under this account, simply reducing the level to $p<.005$, as Benjamin et al. (2018) suggested, and many signees supported, should have the same effect. We accordingly tested this proposed method on the datasets we obtained. Applying the 0.005 criterion reduced the number of originally significant papers to 44, out of which 21 were replicated (47%), similar to the effect of our proposed FDR approach (46%). Critically, however, this criterion led to a loss of 11 replicated results (out of 32), translating into 66% power, which is much worse than the 97% power of the FDR approach. This shows that the heart of the problem lies in selective inference rather than simply the level of evidence. Figure 5 further demonstrates this point; if the problem was indeed the level of evidence, we would expect to find a separation between replicated and non-replicated results, with the latter having closer to threshold p-values. That is, most of the red circles would have been on the far-right end of the x-axis, while most of the blue triangles would have been on the far left of this axis. Instead, we find that the p-values are distributed all over the x-axis for both types of studies. Moreover, the non-replicated ones show a greater difference between the original and the TreeBH p-values, implying again that evident selective inference is a strong factor driving the lack of replicability.

**Discussion**

This work highlighted the importance of adjusting for selective inference, and how it might increase replicability. We showed that "statistically significant" results that were deemed not significant after appropriate correction for multiple comparisons rarely got replicated in the tested database. By adjusting for FDR, replicability can be enhanced without giving up much power. Below we consider the implications of these findings.

**Isn't it the p-value's fault?**

Most researchers use significance testing as a decision tool to separate the highlighted experimental results from those prone to be generated by chance. Therefore, it is natural that much of the blame for the replication crisis pertained to general questionable research practices that are related to the use of p-values, leading to attacks on the p-value itself. The discussion around this point evoked much interest within and outside the statistical sciences community. For example, attacks on the p-value, published in psychological journals (Trafimow & Marks, 2015; Lee, 2016; Amrhein et al., 2019) and conveyed in leading scientific and non-scientific venues, led the American Statistical Association (ASA) to publish guidelines where alternative measures were discussed (Wasserstein & Lazar, 2016). However, later publications offered a broader view of the connection between replicability and statistical practices. This consensus report written in response to then ASA president request (Benjamini et al., 2021), stated that that p-value and significance tests, when properly applied and interpreted, do increase the rigor of the conclusions drawn from the data. They further outline some principles relevant to the use of all statistical methods.

Importantly, alternative statistical methods do not offer an advantage over the p-value in minimizing the selection problem (Savalei & Dunn, 2015; Benjamini, 2020). As for confidence intervals (CIs), their calculation (frequentists/Bayesian) requires the specification of the alpha level; p-values and p-values adjusted for selection do not. Checking whether a 95% CI covers the null is equivalent to the use of p-value $\leq 0.05$ and carries the same problems. More critically, these unadjusted CIs, when restricted to the selected ones, fail to offer the assumed level of coverage. Hence, the proper use of CIs also requires adjustment for selection (Benjamini & Yekutieli, 2005), a rare practice in psychology. Thus, the move toward unadjusted CIs instead of the p-values, suggested in the ASA statement, is irrelevant to the replicability concern . Moreover, since the p-values are the most assumptions-free statistical tool, and these minimal assumptions hold for well-designed experiments, we hold that it should remain a central inference tool in tandem with CIs when the latter are relevant and feasible (Benjamini 2020) .

Using the TreeBH-adjusted alpha of the Level to construct a CI will create an adjusted CI that, when it does not cover the null, will be identical to rejecting the null using the TreeBH p-value. However, it has yet to be proven that a TreeBH-adjusted 95% CI will cover the parameter 95% of the time.

As to Bayesian-based methods of inference, they have the reputation of being immune to selective inference. Unfortunately, the theory backing this reputation relies on the assumption that the studied effects are sampled from an exactly prescribed prior. It has been shown that even a minute deviation from the assumed prior (one in a thousand) may lead to a drastic failure of the selected posterior-based CIs (Benjamini, 2020). Empirical Bayes analyses try to circumvent the problem by estimating the prior in large problems, but this is still followed by addressing the selection problem (Efron, 2015).

**Multiple replications**

The RPP, and hence our study, involved a single replication effort for each result. This approach has been criticized, claiming that multiple replications are favorable over one definitive replication (Schooler, 2014). This translated into multi-lab projects, and multiple replicability attempts (Benjamini & Heller, 2008; Valentine et al., 2011; van Aert & van Assen, 2017, 2018; Hung & Fithian, 2019; Pavlov et al., 2021). We join the call but note that these approaches improve the replication process rather than increase the replicability of new findings; only after all these multi-lab replication attempts have been made can we estimate the replicability of the original finding. Our method, instead, aims at *improving the replicability of new findings even before any replication attempt has been made*. By weeding out findings likely to stem from false discoveries, we strive to minimize the chances of publishing findings that are less likely to be replicated.

## Conclusion

Psychological science has been taking substantial steps to improve replicability, decrease false discoveries, and generate a more reliable corpus of knowledge. In addition to addressing not-evident selective inference, which has already become common practice, we emphasize the need to adjust for evident selective inference. In view of our results, highlighting merely unadjusted "significant" p-values that would not pass the significance threshold once adjusted is misleading and should be avoided. Hence, addressing evident selective inference is an essential missing piece in the replicability puzzle, promising to enhance replicability in experimental psychology. We hope it will be adopted as a minimal standard for good scientific praxis in the field and become common practice when reporting statistical results. Specifically, we find controlling the false discovery to be an appropriate criterion for addressing selective inference in experimental psychology, as it balances the protection from false discoveries and retainment of power. Researchers can be assured that using it should not unduly lower the probability of finding effects when they are real; it would only minimize the proportion of null effects among the highlighted significant ones, hereby increasing replicability.

Whichever adjustment method one uses, it should be specified, and we recommend reporting both the original exact p-value and the adjusted one (akin to how we reported the results above). This would echo similar praxis from other fields where this is a standard requirement by journals and reviewers (e.g., fMRI studies; Nguyen et al., 2014).